\documentclass[fleqn,twoside]{article}
\usepackage{poster2}

\usepackage{graphicx}
\usepackage[figuresright]{rotating}

\newcommand{\AmS}{{\protect\the\textfont2
  A\kern-.1667em\lower.5ex\hbox{M}\kern-.125emS}}

\include{definitions}

\title{$\tau$ Lepton Decays and CVC}

\author{ V.~A.~Cherepanov\thanks{Talk at the X Workshop
on $\tau$ Lepton Physics, Novosibirsk, September 2008}  and S.~I.~Eidelman \\ {Budker Institute of Nuclear Physics, \\ 
        11 Lavrentyev St., \\ 
        630090 Novosibirsk, Russia} }%

\begin{document}

\begin{abstract}
We use experimental data on $e^+e^- \to \eta (\eta^{\prime})\pi^+\pi^-$ and
conservation of vector current to estimate the branching fractions
of $\tau^-$ decay  to $\eta (\eta^{\prime})\pi^-\pi^0\nu_{\tau}$. The
obtained values are compared to the experimental results.
\vspace{1pc}
\end{abstract}

\maketitle

\section{Introduction}

Low energy $e^+e^-$ annihilation into hadrons provides a source of 
valuable information about the interactions of light quarks. Precise
measurements of the exclusive cross sections as well as the 
total cross section appear important for different applications 
like, e.g., determination of various QCD parameters like quark masses,
quark and gluon condensates~\cite{qcd}, 
calculations of the hadronic contributions to the muon
anomalous magnetic moment and running fine structure
constant~\cite{ej}.
 
The hypothesis of conserved vector current (CVC) and 
isospin symmetry relate to each other the isovector part of
$e^{+}e^{-} \rightarrow$ hadrons and corresponding (vector current
induced) hadronic decays of the $\tau$-lepton~\cite{tsai,saku}.
These relations provide a possibility to use an independent 
high-statistics data sample from $\tau$ decays 
for increasing the precision of the
spectral functions directly measured in $e^+e^-$ annihilation~\cite{adh}.
  
While this idea appeared fruitful 10 years ago~\cite{adh}, further
increase of the experimental statistics in both $e^+e^-$ and $\tau$
sectors revealed unexpected problems: the $2\pi$ spectral function 
determined from $\tau$ decays using CVC  was significantly higher than that
obtained from $e^+e^-$, there are indications at a similar deviation
in the four-pion channel~\cite{dehz1,dehz2}.

It is therefore interesting to perform a systematic test of CVC
relations using available experimental information on various 
final states.  For the vector part of the weak hadronic current 
the mass distribution of the  produced hadronic system is

\small
\begin{equation}
\frac{d\Gamma}{dq^{2}} = 
\frac{G_{F}|V_{\rm ud}|^{2}S_{\rm EW}}{32\pi^{2}m^{3}_{\tau}}(m^{2}_{\tau}
- q^{2})^{2}(m^{2}_{\tau}+2q^{2})v_{1}(q^{2}), 
\end{equation}

\normalsize
\noindent
where a spectral function $v_{1}(q^{2})$ is given by the expression

\begin{equation}
v_{1}(q^{2}) =
\frac{q^{2}\sigma^{I=1}_{e^{+}e^{-}}(q^{2})}{4\pi\alpha^{2}},
\end{equation}
\noindent
and $S_{\rm EW}$ is an electroweak
correction equal to 1.0194 according to~\cite{marc}.

The allowed quantum numbers for the hadronic decays channels are:


\begin{equation}
J^{PG} = 1^{-+}, \tau \rightarrow 2n\pi\nu_{\tau}, 
\omega\pi\nu_{\tau}, \eta\pi\pi\nu_{\tau}, \ldots 
\end{equation}

\noindent
After integration
\small
\begin{eqnarray}
\frac{{\cal B}(\tau^{-} \rightarrow X^{-}\nu_{\tau})}
{{\cal B}(\tau^{-} \rightarrow e^{-}\nu_{e}\nu_{\tau})} = 
\frac{3|V_{\rm ud}|^{2}S_{\rm EW}}{2\pi\alpha^{2}}\times \nonumber \\
\int_{4m^{2}_{\pi}}^{m^{2}_{\tau}}dq^{2}
\frac{q^{2}}{m^2_{\tau}}(1-\frac{q^{2}}{m^{2}_{\tau}})^{2}
(1+2\frac{q^{2}}{m^{2}_{\tau}})\sigma^{I=1}_{e^{+}e^{-}}(q^{2}).
 \end{eqnarray}
\normalsize

Theoretical predictions for the branching ratios of 
different $\tau$ decay modes based on CVC were earlier given 
by many authors, see the bibliography in Ref.~\cite{eid}.
New comparison of CVC based predictions with experiments on $\tau$
lepton decays was motivated by recent progress 
of experiments on $\tau$ decays as well as by updated 
information from $e^{+}e^{-}$ annihilation into hadrons, 
coming mostly from the BaBar collaboration. In this work we focus 
on two particular final states -- $\eta \pi^+ \pi^-$ and
$\eta^{\prime} \pi^+ \pi^-$. 

For numerical estimates we will use the value of the electronic
branching
${\cal B}(\tau \rightarrow e\nu_{e}\nu_{\tau})=$ 17.85 $\pm$ 0.05 \%
and $|V_{\rm ud}|^2$=0.9742 
recommended by RPP-2008~\cite{pdg}.


\section{$\tau^- \rightarrow \eta\pi^{-}\pi^{0}\nu_{\tau}$}
\begin{figure}[htb]
\vspace{9pt}
\includegraphics[width=17pc,height=10pc,scale=0.8]{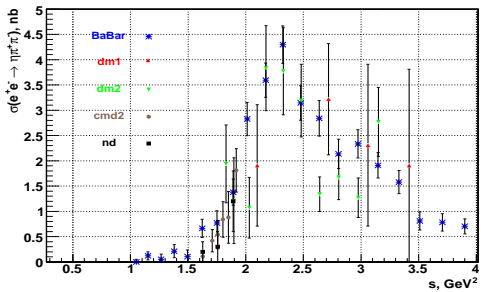}
\caption{Cross section of the process 
$e^{+}e^{-} \rightarrow \eta\pi^{+}\pi^{-}$}
\label{eta}
\end{figure}

The reaction $e^{+}e^{-} \rightarrow \eta\pi^{+}\pi^{-}$ was
recently studied by the BaBar collaboration using ISR in a broad
energy range~\cite{bab}. Earlier
measurements were performed at the ND~\cite{nd}, CMD-2~\cite{cmd2}
detectors from 1.25~GeV to 1.4~GeV and at the DM1~\cite{dm1} and DM2~\cite{dm2}
detectors above this energy. Figure~\ref{eta} shows results of various
measurements. In general, they are in fair agreement with each
other within errors although below 1.4~GeV the values of the cross
section from BaBar are somewhat higher than those of previous
experiments. Above this energy the results of BaBar are significantly
higher than those of DM2, whereas they are in good agreement with
DM1. However, the measurement of the latter has much worse accuracy  
compared to DM2. 

We calculated the branching fraction of $\tau^- \to \eta\pi^-\pi^0\nu_{\tau}$
decay expected from the above mentioned $e^+e^-$ data using the
relation (4).  
The direct integration of experimental points in the energy range 
from 1.25~GeV to the $\tau$ mass using older data 
gives for the branching ratio $(0.132 \pm 0.016)\%$ in agreement with
the previous estimate~\cite{eid}
while the one based on the BaBar data 
gives $(0.165 \pm 0.015)\%$, where we took into account the 8\% systematic 
uncertainty claimed by the authors~\cite{bab}. Averaging and inflating 
the error by a scale factor of 1.50, we arrive at the CVC prediction
of $(0.150 \pm 0.016)\%$ for the energy range
(1.25-1.77)~GeV. Finally,
we add the contribution of the low energy range from 1.0~GeV to 1.25~GeV
(the BaBar data set) to obtain the total CVC expectation of 
$(0.155 \pm 0.017)\%$. It can be compared to the results of the measurements
which are shown in Table~\ref{tabeta} and include 
older results from CLEO~\cite{cleo1} and ALEPH~\cite{aleph} 
as well as the new result of Belle reported at this Workshop~\cite{belle}: 
$(0.135 \pm 0.003 \pm 0.007)\%$.
\begin{table*}[htb]

\hspace{81pt}Table 2

\hspace{81pt}Experimental values of ${\cal B}(\tau^- \to \eta\pi^-\pi^0\nu_{\tau})$

\begin{center}
\label{tabeta}
\newcommand{\m}{\hphantom{$-$}}
\newcommand{\cc}[1]{\multicolumn{1}{c}{#1}}
\renewcommand{\tabcolsep}{2pc} 
\renewcommand{\arraystretch}{1.2} 
\begin{tabular}{@{}lll}
\hline
Group &  ${\cal B}$, \% & Ref. \\
\hline
CLEO, 1992   & $0.170 \pm 0.020 \pm 0.020$   & \cite{cleo1} \\
ALEPH, 1997 &$0.180 \pm 0.040 \pm 0.020$  & \cite{aleph} \\
Belle, 2008 & $0.135 \pm 0.003 \pm 0.007$ & \cite{belle} \\ 
\hline
\end{tabular}\\[2pt]
\end{center}
\end{table*}

The average of the experimental results gives 
 ${\cal B}(\tau^- \to \eta\pi^-\pi^0\nu_{\tau})=(0.139 \pm 0.008)\%$,
which is $0.9\sigma$ lower than the prediction above.
 
It is also interesting to compare our result with earlier
theoretical estimates of this branching fraction, see
Table~\ref{tabetap}.  
\begin{table*}[htb]

\hspace{81pt}Table 2

\hspace{81pt}Theoretical predictions for  ${\cal B}(\tau^- \to \eta\pi^-\pi^0\nu_{\tau})$
\label{tabetap}
\newcommand{\m}{\hphantom{$-$}}
\newcommand{\cc}[1]{\multicolumn{1}{c}{#1}}
\renewcommand{\tabcolsep}{2.45pc} 
\renewcommand{\arraystretch}{1.2} 
\begin{center}
\begin{tabular}{@{}lll} 
\hline
Method & ${\cal B}, \%$ & Ref. \\
\hline
$\rho^{\prime}$  & $\sim 0.3$ & \cite{pich87} \\
CVC  & $\sim 0.15$ & \cite{gilman} \\
Eff. Lagr.  & $0.14^{+0.19}_{-0.10}$ & \cite{eric} \\
Eff. Lagr.  & 0.18--0.88 & \cite{kram} \\
CVC   & $0.13 \pm 0.02$ & ~\cite{eid} \\ 
CVC  & $0.14 \pm 0.05$ & \cite{nari} \\
CVC + Eff. Lagr.   & $\sim 0.19$ & \cite{deck} \\
Eff. Lagr.   & $\sim 0.19$ & \cite{li} \\
\hline
\end{tabular}\\[2pt]
\end{center}
\end{table*}
It can be seen that the older predictions based on the $e^+e^-$ data 
and CVC agree with the much more accurate result of this work, 
which uses more data,
in particular a more precise data sample of BaBar. Other predictions,
which are more theoretically driven and use the low-energy effective
Lagrangian, show a much larger spread of the results.

\section{$\tau^- \rightarrow \eta^{\prime}\pi^{-}\pi^{0}\nu_{\tau}$}
\begin{figure}[htb]
\vspace{7pt}
\includegraphics[width=17pc,height=10pc,scale=0.6]{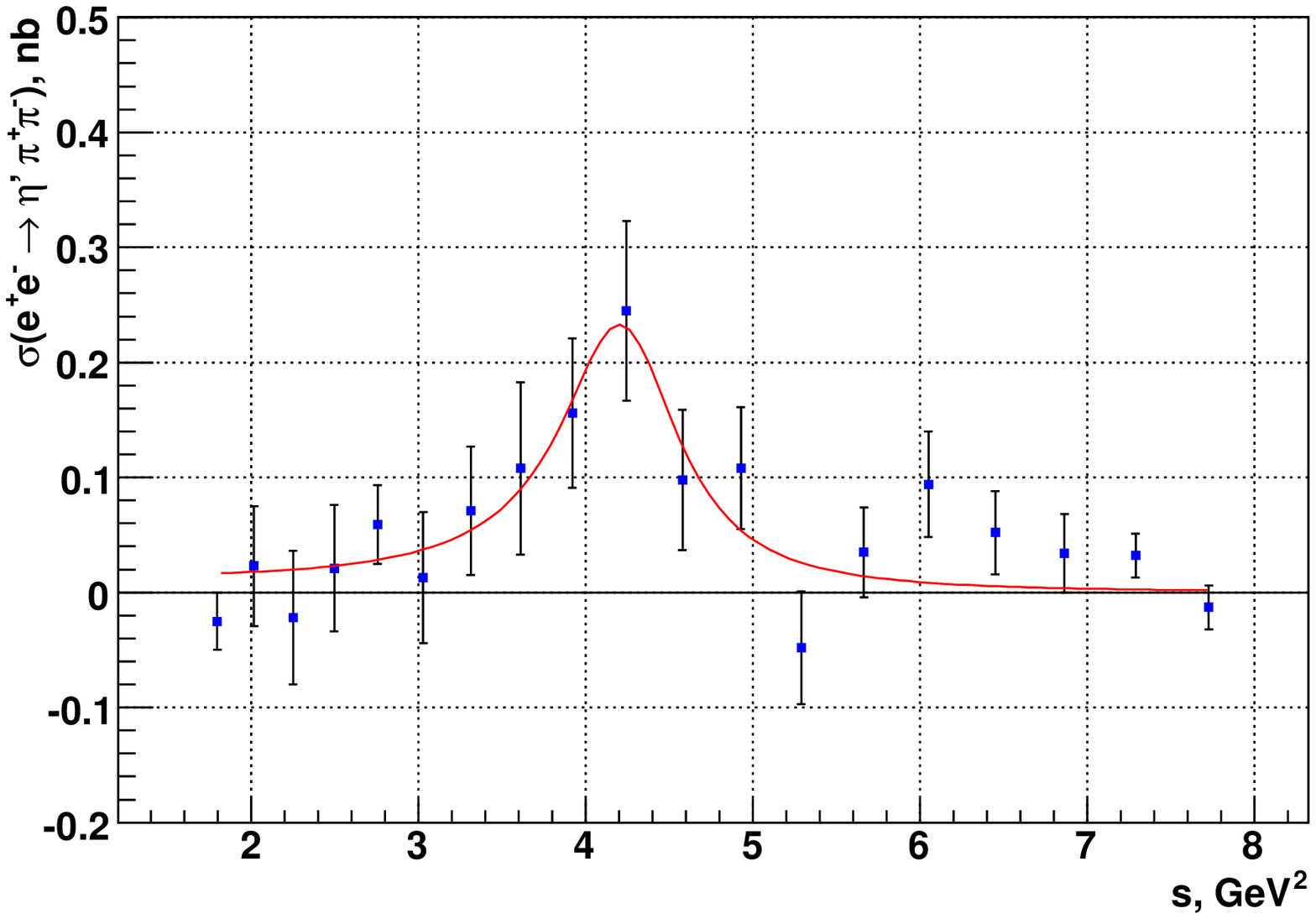}
\caption {Cross section of the process
$e^{+}e^{-} \rightarrow \eta^{\prime}\pi^{+}\pi^{-}$}
\label{fig:etap}
\end{figure}

Recently, the BaBar collaboration presented the very first measurement
of the cross section of the process
$e^{+}e^{-} \rightarrow\eta^{\prime}\pi^{+}\pi^{-}$~\cite{bab},
see Fig.~\ref{fig:etap}. The cross section of the process clearly shows
resonant behavior with a maximum slightly above 2~GeV, but its values
at energies below the $\tau$ lepton mass are very small. We fit the
cross section assuming the Breit-Wigner amplitude for production of
three pseudoscalar mesons~\cite{ps3} and obtain
the following resonance parameters (mass, width and cross section at
the peak): 

\begin{equation}
M = 2071 \pm 32~{\rm MeV},~~
\end{equation} 
\begin{equation}
\Gamma = 214 \pm 76~{\rm MeV},~~
\end{equation}
\begin{equation}
\sigma_0 = 0.223 \pm 0.073 \pm 0.022~{\rm nb}. 
\end{equation}

Here the systematic error of $\sigma_0$ is 10\% following the estimate
of the overall systematic error of the cross section in Ref.~\cite{bab}. 
Taking into account that we do not estimate systematic uncertainties
on mass and width one can conclude that the resonance parameters are
compatible with those of the $\rho(2150)$~\cite{pdg}. Using (4) we 
integrate the optimal curve for the cross section up to the $\tau$
mass and obtain for the branching ratio: 
\small
\begin{equation}
{\cal B}(\tau^{-} \rightarrow \eta^{\prime}\pi^{-}\pi^{0}\nu_{\tau})=
(13.4 \pm 9.4 \pm 1.3 \pm 6.1) \times 10^{-6},
\end{equation}
\normalsize
where the first error is statistical (that of the fit), the second is
experimental systematic and the third is the model one estimated by
using the world average values of the $\rho(2150)$ mass and width
and varying them within errors. The obtained result is consistent with
zero and we place the following upper limit at 90\% CL using the
method of Ref.~\cite{fc}: 
\begin{equation}
{\cal B}(\tau^{-} \rightarrow \eta^{\prime}\pi^{-}\pi^{0}\nu_{\tau})<
3.2 \times 10^{-5},
\end{equation}
which is two times more restrictive than the upper limit based on the
only existing measurement from CLEO~\cite{cleo2}:
\begin{equation}
{\cal B}(\tau^{-} \rightarrow \eta\prime\pi^{+}\pi^{-}\nu_{\tau})< 
8 \times 10^{-5},
\end{equation}
but still an order of magnitude higher than a theoretical estimate
${\cal B}(\tau^{-} \rightarrow \eta^{\prime}\pi^{-}\pi^{0}\nu_{\tau}) 
\approx 4.4 \times 10^{-6}$ based on the chiral Lagrangian~\cite{li}.

\section{Conclusions}
Using available data from $e^+e^-$ annihilation and CVC we obtained 
the following results for the $\tau$ lepton branching fractions: 
\begin{itemize}
\item 
for the $\eta \pi^-\pi^0\nu_{\tau}$ the expected branching
is $(0.155 \pm 0.017)\%$ compatible with the world average of 
$(0.139 \pm 0.008)\%$; 
\item
for the  $\eta^{\prime} \pi^-\pi^0\nu_{\tau}$ the upper limit is
$< 3.2 \times 10^{-5}$ or 2.5 times smaller than the
experimental upper limit of $< 8 \times 10^{-5}$, both at 90\% CL.  
\end{itemize}

We are grateful to D.A.~Epifanov and E.P.~Solodov for useful
discussions. This work was supported in part by grants RFBR 06-02-16156, 
RFBR 08-02-13516, RFBR 08-02-91969, INTAS/05-1000008-8328, PST.CLG.980342
and DFG GZ RUS 113/769/0-2.

\end{document}